\documentclass{moriond}

\usepackage{amssymb,amsmath}
\def\be{\begin{equation}}
\def\ee{\end{equation}}
\def\bea{\begin{eqnarray}}
\def\eea{\end{eqnarray}}

\def\a{\alpha}

\def\g{\gamma}

\def\r{\rho}

\def\D{\Delta}

\def\O{\Omega}

\newcommand{\vev}[1]{\langle #1 \rangle}

\def\be{\begin{equation}}
\def\ee{\end{equation}}

\def\bea{\begin{eqnarray}}
\def\eea{\end{eqnarray}}

\newcommand{\Photo}{\vspace*{2mm}\includegraphics[width=0.25\linewidth]{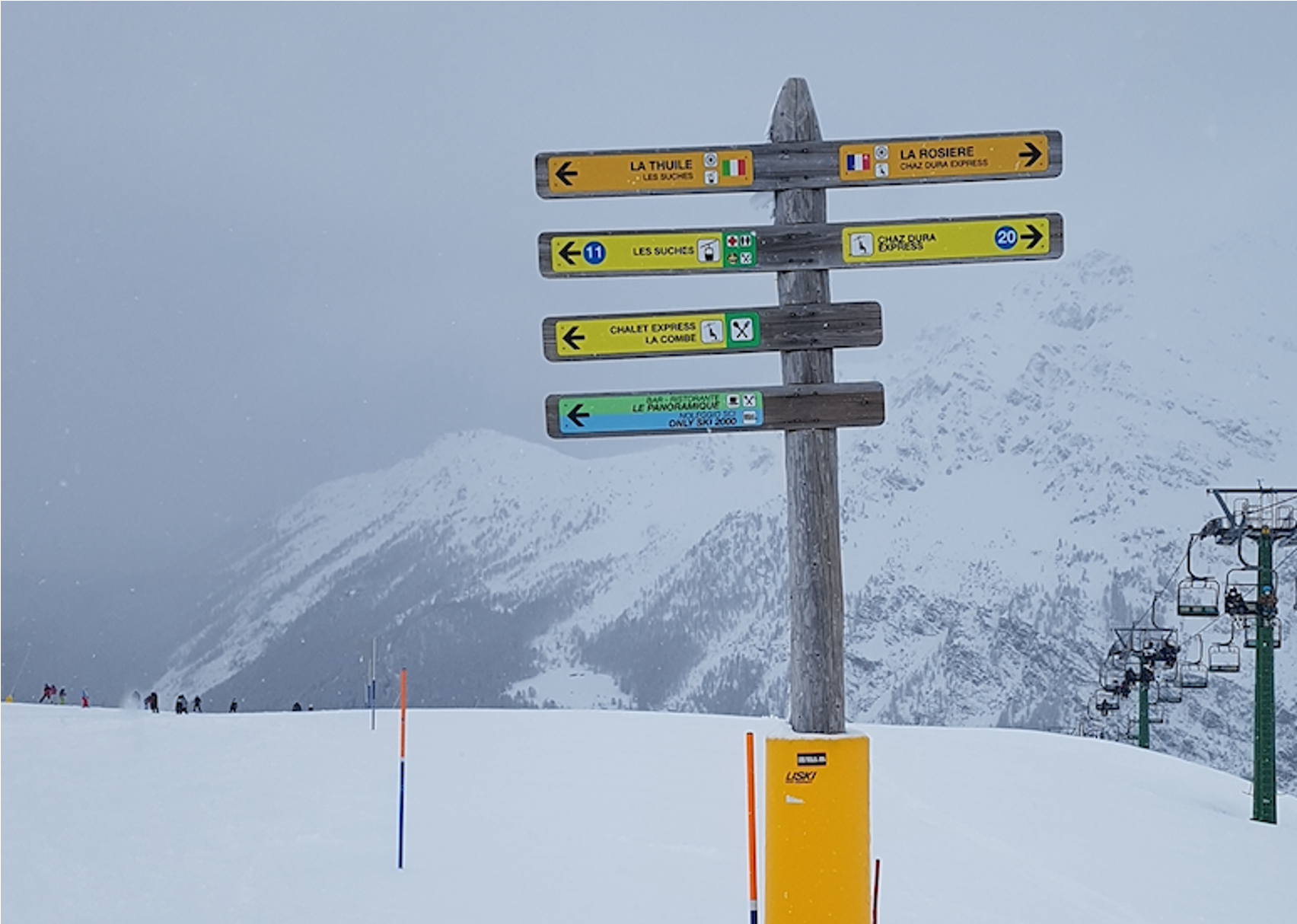}}

\begin{document}
\vspace*{2cm}
\title{Majorana mass generation, gravitational waves and cosmological tensions\footnote{Talk based on \cite{paper1,paper2}.}}

\author{Pasquale Di Bari}

\address{School of Physics and Astronomy, University of Southampton, Southampton, SO17 1BJ, U.K.}

\maketitle\abstracts{
A neutrino Majorana mass generation in the early universe 
might have left imprints in cosmological observables.  It can
source the production of a detectable stochastic background of primordial gravitational waves
with a spectrum that combines a contribution from a first order phase transition production 
and from the vibration of global cosmic strings. An intriguing possibility is given by 
the split seesaw model. In this case, in addition to the traditional
high scale seesaw, a low scale neutrino Majorana mass generation can solve a potential primordial deuterium problem
and ameliorate the cosmological tensions of the $\Lambda$CDM model.
At the same time, it can also produce subdominant contribution to the NANOGrav signal of a 
stochastic background of gravitational waves (GWs), 
in addition to the astrophysical dominant contribution from supermassive black hole mergers.}

\section{A neutrino solution to the problem of the origin of matter in the universe}

Despite there is no evidence of new physics at colliders (so far) \cite{expsummary}, 
the necessity to explain neutrino masses and the origin of matter in the  universe 
requires the existence of new physics.
Minimal WIMP dark matter models (those minimal ones realising the WIMP miracle) and electroweak baryogenesis have for long time been regarded as the natural solution to the problems of dark matter and matter-antimatter asymmetry of the universe, respectively. They are   nicely  
both consistent with the idea of new physics close to the electroweak scale, as independently suggested by naturalness. 
However, the null results at colliders 
and the strong constraints
on WIMPs imposed by direct and indirect searches,  have in the last twenty years stimulated 
the rise of many new  ideas and solutions beyond a `natural' solution.

In particular, the discovery of neutrino masses  in neutrino oscillation experiments make quite attractive those
solutions that rely on extensions of the standard model (SM) able to explain neutrino masses. The most popular one is certainly represented by a minimal type-I seesaw extension of the SM, introducing of $N \geq 2$ 
right-handed (RH) neutrinos, with Yukawa couplings to the left-handed lepton doublets and Higgs doublet and with a right-right Majorana mass term.          
This simple extension allows, notoriously, to explain the matter-antimatter asymmetry of the universe with leptogenesis \cite{fy}
and even a dark matter solution where the lightest RH neutrino, with a keV mass, plays the role of dark matter
\cite{dw,asaka}.  Within this simple bottom-up approach, while the Dirac mass term is generated at the electroweak spontaneous symmetry breaking, the Majorana mass is a given term in the fundamental lagrangian. However, it is
reasonable that the Majorana mass term is also generated dynamically, in one or even multiple stages.   

\section{Majorana mass generation in the majoron model}

There are different ways to describe the generation of a Majorana mass term. In a traditional top-down approach 
a Majorana mass term generation is described in grandunified models such as $SO(10)$ and it is 
related to the generation of all other mass terms. A simple bottom-up approach is to consider a majoron model \cite{majoron}.
Introducing a complex scalar field $\phi =({\varphi / \sqrt{2}})\, e^{i\theta}$, coupling to the RH neutrinos, one has
the following simple extension of the SM Lagrangian,
\bea\label{eq:L_l}
-{\cal L}_ {N_I+\phi} & = & 
\textcolor{black}{\left( \overline{L_{\alpha}}\,h_{\alpha I}\, N_{I}\, \widetilde{\Phi} 
+  {\lambda_{I}\over 2}  \, \phi \, \overline{N_{I}^c} \, N_{I}
+ {\rm h.c.}\right) + V_0(\phi) \,,}
\eea
where $V_0(\phi)$ is the tree level potential associated to $\phi$. 
At sufficiently high temperatures thermal effects enforce $\langle\phi\rangle = 0$
in a way to have global $U(1)_L$ symmetry restoration. Below a certain temperature $T_{\star}$
the  complex scalar field $\phi$ undergoes a phase transition at the end of which 
$\langle \phi \rangle = e^{i\,\theta_0}\,v_0/\sqrt{2}$ and $U(1)_L$
is spontaneously broken. This generates a Majorana neutrino mass term, so that RH
neutrinos become massive with masses $M_I = \lambda_I\,v_0/\sqrt{2}$.   
After spontaneous symmetry breaking the field can be written as
\be
\phi= {e^{i\theta_0}\over\sqrt{2}}\,(v_0 +S) e^{i{J\over v_0}} \simeq {e^{i\theta_0} \over \sqrt{2}}\,(v_0 +  S + i \, J) \,  ,
\ee
 where $S$ is a massive real scalar describing quantum fluctuations in the radial direction ($S = \delta \varphi$)
 and $J$ is the majoron, a massless Goldstone field (an example of axion-like particle), describing angular quantum
 fluctuations ($J = v_0\,\delta \theta$).
 Finally, after electroweak symmetry breaking, a Dirac mass term $m_D = v_{\rm ew} \, h$ is also generated and 
a light neutrino mass matrix given by the seesaw formula
$(m_\nu)_{\alpha\beta} = - m_D \, M^{-1}\,  m_D^T$,
where $v_{\rm ew}$ is the SM Higgs vev.
We will refer to the SM particles as the visible sector while to the RH neutrinos $N_I$, the majoron $J$ and massive boson $S$ as the dark sector.  For definiteness, we have assumed $T_\star$ higher than the temperature of the electroweak symmetry phase transition but  the case when Majorana mass generation occurs below the electroweak scale is also possible, we will also consider this option.   

\section{Gravitational waves from Majorana mass generation}

The calculation of the GW spectrum, starting from the lagrangian, proceeds through a few steps. 
First of all one needs to calculate the {\em dressed effective potential} including thermal effects at 1 loop and resummed thermal
masses that account for higher order effects.  From a high temperature expansion, this can be written in the general 
polinomial form
\be\label{VTeffminimal2}
 V^T_{\rm eff}(\varphi) \simeq D\,(T^2 - T_0^2) \, \varphi^2 - (A \, T + \widetilde{\mu}) \, \varphi^3 + \frac{1}{4}\lambda_T\, \varphi^4 \,  ,
\ee
where the different coefficients can be expressed in terms of the parameters of the lagrangian. 
An important role is played by $\widetilde{\mu}$, since its presence creates a barrier at zero temperature and this makes the transition
much stronger greatly enhancing the GW production.   From the effective potential one can calculate the euclidean action $S_E(T)$
that gives the bubble nucleation rate.  

From the euclidean action one can derive different quantities characterising the phase transition and the GW spectrum. 
First of all, one can calculate the {\em percolation temperature} that can be identified with the temperature of the phase transition if
its duration is short enough, an approximation valid for not too strong phase transitions. Such duration is described by the quantity $\beta/H_\star$, giving 
approximately the age of the universe-to-phase transition duration at the temperature of the phase transition: the higher is $\beta/H_\star$ the shorter is the duration of the phase transition compared to the age of the universe and vice-versa.  
One can also calculate the 
strength of the phase transition parameter $\alpha \equiv \varepsilon(T_\star)/\rho_{\rm R}(T_\star)$, where 
$\varepsilon$ is the latent heat feed in the phase transition and $\rho_{\rm R}$ is the total energy density of radiation. 
If the dark sector is coupled to the visible sector, then its temperature $T_{\rm D} = T$, otherwise one has to distinguish
the two temperatures  and also calculate separately $\alpha_{\rm D} = \varepsilon(T_{\rm D}^\star)/\rho_{\rm RD}(T_{\rm D}^\star)$ \cite{Breitbach:2018ddu,Fairbairn:2019xog,Bringmann:2023opz},
the strength of the phase transition in the dark sector. From these parameters one can calculate the GW spectrum from 
first order phase transition due to the nucleation of bubbles. The GW spectrum is  defined as
\be
h^2\O_{{\rm GW}0}(f)=  {1 \over \rho_{{\rm c}0}h^{-2}} \,  {d\rho_{{\rm GW}0}\over d\ln f} \,  ,
\ee 
where $\rho_{{\rm c}0}$ is the critical energy density  and $\rho_{{\rm GW}0}$ is
the  energy density of GW, produced during the phase transition, both calculated at the present time. 
This is given by the sum of three contributions:  one from turbulence, one from bubble collisions and one from
sound waves \cite{Caprini:2015zlo}. 
From current calculations, the sound wave contribution certainly dominates for $\alpha \lesssim 0.3$. In this
case one has quite a reliable semi-analytical expression derived within a sound shell model \cite{Hindmarsh:2016lnk},
and supported by numerical calculations \cite{Hindmarsh:2017gnf}, given by
\be\label{omegasw}
h^2\Omega_{\rm sw 0}(f) =3 \, h^2 \, r_{\rm gw}(t_\star,t_0)\,\widetilde{\Omega}_{\rm gw} \,  
{(8\pi)^{1/3}\,v_{\rm w}\over \beta/H_\star} \, \left[\frac{\kappa(\alpha)\, \alpha}{1+\alpha}\right]^2 \, 
\widetilde{S}_{\rm sw} (f) \, \Upsilon(\alpha,\beta/H_{\star})\, ,
\ee
where $v_{\rm w}$ is the bubble wall velocity,   $\kappa(\alpha)$ is an efficiency factor giving the fraction of energy transferred
in GWs, $\Upsilon(\alpha,\beta/H_{\star})$ is a suppression factor accounting for the duration of the GW production \cite{Guo:2020grp}.  
The normalised spectral shape function is given by $\widetilde{S}_{\rm sw} (f) \simeq 0.687\, S_{\rm sw} (f)$ with
\begin{eqnarray}\label{Ssw}
S_{\rm sw} (f) = \left(\frac{f}{f_{\rm sw}}\right)^3 \left[\frac{7}{4+3({f/f_{\rm sw}})^2} \right]^{7/2} \,  ,
\end{eqnarray} 
where $f_{\rm sw}$ is the peak frequency given by
\begin{eqnarray} \label{fpeak}
f_{\rm sw} =8.9\,\mu{\rm Hz} \, \frac{1}{v_{\rm w}} \frac{\beta}{H_\star} \left( \frac{T_\star}{\rm 100\,GeV}\right) \left( \frac{g_{\rho\star}}{106.75} \right)^{1/6} \, .
\end{eqnarray}
Finally,  the redshift factor $r_{\rm gw}(t_\star,t_0) $, evolving $\Omega_{\rm gw\star} \equiv \rho_{\rm gw\star}/\rho_{{\rm c}\star}$ into $\Omega_{\rm gw 0} \equiv \rho_{\rm gw 0}/\rho_{\rm c 0}$, is  given by \cite{Kamionkowski:1993fg}
\be
r_{\rm gw}(t_\star,t_0) = \left({a_\star \over a_0}\right)^4 \, \left({H_\star \over H_0}\right)^2 
= \left({g_{S0}\over g_{S\star}}\right)^{4 \over 3}\,{g_{\r\star} \over g_\gamma} \, \O_{\gamma 0}
 \simeq 3.5 \times 10^{-5} \, \left({106.75 \over g_{\rho \star}} \right)^{1\over 3} \,  \left({0.6875 \over h}\right)^2 .
\ee
The peak frequency depends linearly on the temperature of the phase transition and, therefore, the experimental identification  
of such a peak would provide a clear indication of the scale of new physics.
For values $\alpha \gtrsim 0.3$, one can expect strong deviations from Eq.~(\ref{omegasw}). Recently, some numerical calculations of $\Omega_{\rm GW,0}$,  the total  GW contribution to the energy density parameter, have been
presented for $\alpha \leq 0.6$ \cite{Cutting:2019zws}.
They show that there is some suppression that is particularly strong (up to three orders of magnitude) 
in the deflagration case, for $v_{\rm w} < c_{\rm s}$.
In the case of detonation, for $v_{\rm w} > c_{\rm s}$, such suppression is contained within one order of magnitude and the suppression is weaker and weaker for higher values of $v_{\rm w}$.
For phase transitions in a dark sector the detonation case holds quite reliably.

As a {\em minimal model} one can consider the usual tree level quartic potential
 \be\label{minimal}
V_0(\varphi) = -{1\over 2}\,\mu^2 \, \varphi^2 + {\lambda\over 4}\,\varphi^4 \,  .
\ee
In this case one has simply $v_0 \equiv \sqrt{\mu^2/\lambda}$, $m_S^2= 2 \lambda v_0^2$,  the majoron $J$ is massless
and, importantly, $\widetilde{\mu} =0$. In this case the GW production turns out to be a few orders of magnitude
below the experimental sensitivity of any experiment \cite{DiBari:2021dri}. 

However, one should also consider a contribution to the GW spectrum from the generation of a global string network 
during the phase transition \cite{paper2}.  Compared to the Nambu-Goto string-induced almost flat gravitational wave spectrum associated with a gauged symmetry breaking, the contribution from global cosmic string-induced is typically suppressed
and one can show that it is $\propto v_0^4$. 
This makes their detection at interferometers more challenging, unless the symmetry breaking scale $v_0$ is above $10^{14}$ GeV. 

\section{Majorana mass generation  and GW production in multiple majoron models}

It was already noticed in the case of the electroweak phase transition \cite{profumo} that the 
addition of an auxiliary scalar field $\eta$ coupling to $\varphi$ can generate a zero temperature
barrier, corresponding to $\widetilde{\mu}\neq 0$, greatly enhancing the GW production. 

The minimal potential in Eq.~(\ref{minimal}) can be then generalised including the contribution from 
an auxiliary (real) scalar field $\eta$ \cite{DiBari:2021dri}: 
 \be
V_0(\varphi,\eta) = V_0(\varphi) + \zeta\,\varphi^2\,\eta^2 -{1\over 2}\,\mu_{\eta}^2\,\eta^2 + {1\over 4}\,\lambda_{\eta}\,\eta^4  \,  .
\ee
If we assume that the scalar field undergoes also a phase transition to a much higher temperature than $T_\star$, settling to its
true vacuum prior to the phase transition of $\varphi$ with a vev $v_\eta \gg v_{\varphi}$, then at the end of the 
phase transition of $\eta$ the effective potential for $\varphi$ can be written again in the polynomial form  Eq.~(\ref{VTeffminimal2})
but this time with $\widetilde{\mu} = \zeta^2 \, v_0 /(4 \lambda_\eta) \neq 0$. 
This greatly enhances the strength of the first order phase transition and, therefore, the GW production that 
can in this case be detectable at future gravitational antennas \cite{paper2}. It is interesting that 
if the scale of the Majorana mass generation, that can be identified with $v_0$, is lowered at the GeV scale,
therefore in this case below the electroweak scale so that Majorana mass is generated after the Dirac mass term, 
then the GW spectrum peaks at mHZ frequencies, within LISA sensitivity. In this way there is an interesting interplay 
with collider searches of GeV RH neutrinos. 

The general idea of introducing an auxiliary field can be realised, specifically, within a multiple majoron model \cite{paper1}. 
We can start first considering a two-majoron model.
In this case one introduces two complex scalar fields denoted by $\phi_1$ and $\phi_3$ with their respective global lepton number symmetries  $U(1)_{L_{1}}$ and $U(1)_{L_3}$. The lagrangian can be written as ($I=1,2,3$)
\be
-{\cal L}_ {N_I+\phi_1+\phi_3}  =  
 \left(\overline{L_{\a}}\,h_{\a I}\, N_{I}\, \widetilde{\Phi} 
+  {y_{1}\over 2}  \, \phi_1 \, \overline{N_{1}^c} \, N_{1} +  {y_{2}\over 2}  \, \phi_1 \, 
\overline{N_{2}^c} \, N_{2} + {y_{3}\over 2}  \, \phi_3 \,\overline{N_{3}^c} \, N_{3}
+ {\rm h.c.}\right)  + V_0(\phi_1, \phi_3)\, .
\ee
As one can see, $\phi_3$ couples only to the RH neutrino $N_3$, whereas $\phi_1$ couples to both $N_1$ and $N_2$. 
This can be ensured by giving nonzero $U(1)_{L_{1}}$ charges to $N_1$ and $N_2$ and half of their complementary charge to $\phi_1$, whereas $N_3$ and $\phi_3$ have similar complementary charges under $U(1)_{L_3}$ only. Furthermore, we have chosen a basis where $\phi_1$ and $\phi_3$ only couple to the diagonal elements of the RH neutrino mass matrix.
As in the single majoron model, we can introduce the radial components of the complex fields, writing 
$\phi_1 = \varphi_1 e^{i \theta_1}/\sqrt{2}$ and $\phi_3 = \varphi_3 e^{i \theta_3}/\sqrt{2}$. 
Again we can assume that the vacuum expectation values are along the real axis, $\vev{\phi_1} = v_1/ \sqrt{2}$ and $\vev{\phi_3} = v_3/ \sqrt{2}$. 
After spontaneous breaking of both $U(1)$ symmetries,  one has this time two majorons
$J_1 = v_1\delta\theta_1$ and $J_3 = v_3\delta\theta_3$. We further assume the hierarchy $v_3 \gg v_1$, so that the RH neutrino mass spectrum is hierarchical $M_3 \gg M_1 \simeq M_2 \simeq M$.
 
The $U(1)_{L_1} \times U(1)_{L_3}$ symmetry allows the usual quadratic and quartic terms for both $\phi_1$ and $\phi_2$
but also a quartic mixing between the two scalars, so that the tree level potential can now be written as
\begin{align}
	V_0(\phi_1, \phi_3) = -{\mu_1^2}|\phi_1|^2 + \lambda_1 |\phi_1|^4 -{\mu_3^2}|\phi_3|^2 + \lambda_3 |\phi_3|^4  + \zeta |\phi_1|^2 |\phi_3|^2 \, . \label{Vzero2}
\end{align}
This quartic mixing is now exactly the kind of term we have seen to be able to generate a term $\widetilde{\mu} \neq 0$
in the effective potential, as we have seen in the generic case of adding an auxiliary scalar field. 
The role of auxiliary field is now played by $\varphi_3$.  This time
one has a first high scale phase transition at a temperature $T_{\star,3} \sim v_3$ where $U(1)_{L_3}$ is broken
and a Majorana mass $M_3 = y_3\,v_3$ is generated and then  a low scale phase transition occurring at
$T_{\star,1} \sim v_1$ where $U(1)_{L_1}$ is broken and Majorana masses $M_1 = y_1\,v_1$ and $M_2 = y_2\,v_1$ are generated.
In this way, we obtain GW spectra that at the peak are within the sensitivity of future GW antennas. 
However, in addition to  the GW contribution from sound waves generated by bubble nucleation, one can also
have an additional contribution from the network of cosmic strings generated by the $U(1)_{L_3}$ symmetry breaking 
that is sizeable if $v_3 \gtrsim 10^{14}\,{\rm GeV}$.

One can of course make a step further and consider a three-majoron model where each majoron is the leftover
of a different $U(1)_{L_I}$ ($I=1,2,3$) symmetry breaking. One introduces now three comples scalar fields $\phi_I$
and extends the SM with the lagrangian
 \be
- L_{N_I+\phi_I} =  \left(\overline{L_a}h_{a I} H N_I + \frac{y_{1}}{2}\phi_1 \overline{N^c_1} N_1 + \frac{y_{2}}{2}\phi_2 \overline{N^c_2} N_2 + \frac{y_{3}}{2}\phi_3 \overline{N^c_3} N_3 +\text{h.c.} \right)
 + V_0(\phi_1, \phi_2, \phi_3),
\ee
imposing a $U(1)_{L_1} \times U(1)_{L_2} \times U(1)_{L_3}$ symmetry. 
Denoting the vevs by $\vev{\phi_I} \equiv v_I$ and  assuming $v_3 \gg v_2 \gg v_1$, the tree-level scalar potential is given by
\begin{align}\label{V3scalar}
V_0 (\phi_1, \phi_2, \phi_3) &= \sum_{I=1,2,3} \left[-\mu_I^2 \phi_I^* \phi_I + \lambda_I (\phi_I^* \phi_I)^2\right] + \sum_{I,J,I\neq J}^{1,2,3} \frac{\zeta_{IJ}}{2} (\phi_I^* \phi_I)(\phi_J^* \phi_J) \,  .
\end{align}
This time there will be a first very high scale phase transition occurring at $T_{\star 3} \sim v_3$ where 
 $U(1)_{L_3}$ is broken and a Majorana mass $M_3 = y_3\,v_3$ is generated. There will be a sizeable GW production from global cosmic strings
 if $v_3 \gtrsim 10^{14}\,{\rm GeV}$. This will be followed by a second phase transition at $T_{\star 2}\sim v_2$
 where  $U(1)_{L_2}$ is broken and a Majorana mass $M_2 = y_2\,v_2$ is generated. There will be also an enhanced 
 GW production at bubble nucleation from sound waves with a peak. Finally, a third phase transition occurs 
 at  $T_{\star 1}\sim v_1$ associated to $U(1)_{L_1}$ symmetry breaking with the generation of a Majorana mass $M_1 = y_1 \, v_1$
 and again a GW production from sound waves with a second peak at lower frequencies. In this way the GW spectrum would be
 certainly quite distinctive with two bumps standing from the smooth spectrum generated by 
 global cosmic strings if $v_3 \gtrsim 10^{14}\,{\rm GeV}$. Finally, we should stress that in all these considerations
 there is the assumption for the reheat temperature to be sufficiently large that the phase transitions can actually occur,
 this is a point that is often missed. 
  
\section{Split majoron model, cosmological tensions and NANOGrav signal}

We have so far considered the case of high scales for the Majorana mass generation, much higher than $\sim 100\,{\rm MeV}$,
that means before the quark-hadron phase transition. In this way the massless majorons, that get thermalised at the phase transition,
contribute in a negligible way to the excess radiation compared to the standard model case and one has not to worry about cosmological
constraints. Let us now consider a setup that we will refer to as {\em split majoron model}.  This is depicted in the figure. 
\begin{figure}
\centerline{\includegraphics[width=0.55\linewidth]{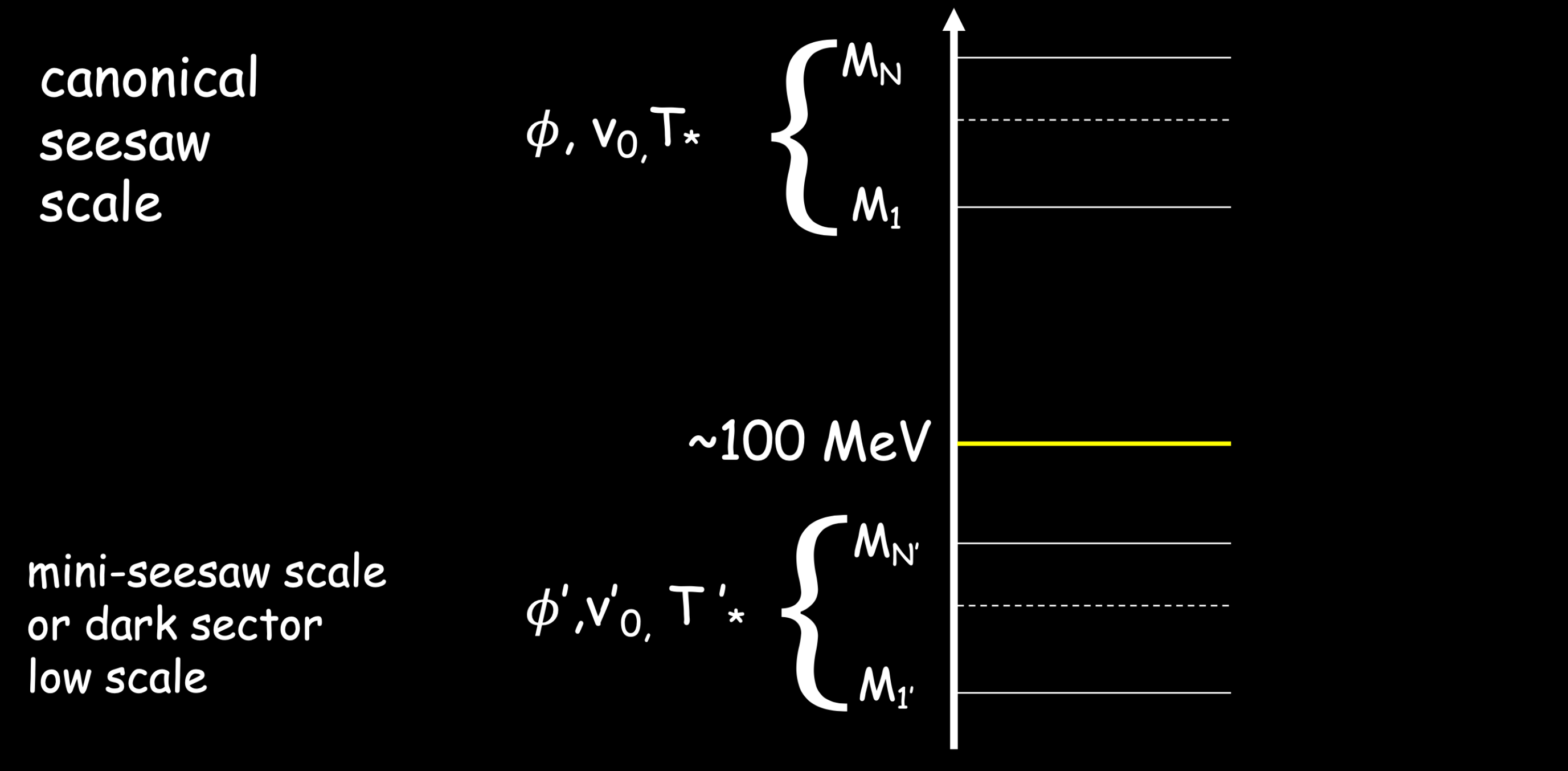}}
\caption{Split seesaw model.}
\end{figure}
As one can see, we assume that in addition to a traditional seesaw scale(s) with $N$ RH neutrinos, 
there is a mini-seesaw scale or a dark sector scale,  below $\sim 100\,{\rm MeV}$, where there are additional $N'$
RH neutrinos and a complex scalar field  $\phi'$. For definiteness we can consider $N=2$ and $N' =1$.  For example,
the lightest RH neutrino could be responsible for the lightest neutrinos mass, as in the $\nu$MSM model \cite{asaka}.

The high scale Majorana masses are generated by the phase transition of a complex scalar field $\phi$.
At the end of the phase transition a massless majoron $J$ is left over but again its contribution to radiation is
strongly diluted by photon production from all SM particle annihilations and can be neglected.
At some temperature $T_\star'$, the complex scalar field $\phi'$ undergoes a first order phase transition 
after which  $U(1)_{L'}$ symmetry is broken a low scale Majorana mass $M'$ is 
generated and a massless majoron $J'$ survives as a cosmological relic. This time its contribution
to extra radiation is not negligible. If one considers a phase transition with $T_\star ' \simeq 10\,{\rm MeV}$, before elecrtroweak
interactions decouple, then the extra radiation contribution from $J'$ parameterised in terms of extra-number of effective neutrinos,
is given by $\D N_{\nu} = 4/7 \simeq 0.6$. A model producing such a fractional amount of extra neutrinos had been proposed, 
after first {\em Planck} satellite results in 2013, as a way to reconcile the Hubble tension, either produced from the decays of a
massive particle \cite{DiBari:2013dna} or just in the form of a Goldstone boson leftover of some $U(1)$ symmetry breaking \cite{Weinberg:2013kea}, correspondingly exactly to our case. 

However, today such a mere injection of extra radiation would reconcile the Hubble tension but it would deform
the positions of CMB peaks in an unacceptable way and something more sofisticate needs to be done. Let us consider a case
where $T_{\star}' \lesssim T_{\nu}^{\rm dec} \sim 1\,{\rm MeV}$. Moreover let us assume that the dark sector has decoupled at 
high energies so that this time $T_{\rm D} \neq T$.  At temperatures below neutrino decoupling, and prior to the low scale phase transition, 
ordinary neutrinos interact with majorons $J$ and complex scalar field $\phi'$ via the effective lagrangian
 \be\label{nudark}
-{\cal L}_{\nu+{\rm D}} = {i\over 2}\,\sum_{i=2,3}\widetilde{\lambda}_i \, \overline{\nu_{i}} \, \g^5 \,\nu_{i} \, J + \zeta\, J\,|\phi'|^2 \,  .
\ee
These interactions can thermalise the ordinary neutrinos to the dark sector after neutrino decoupling ({\em rethermalisation}), in a way that they reach a common temperature
\be
T_{\nu +{\rm D}} = T_\nu^{\rm SM}(T) \, \left({N_{\nu}^{\rm SM}(T) \over N_{\nu}^{\rm SM}(T) + N' + 12/7 + 4\,\D g/7} \right)^{1 \over 4} \,  .
\ee
For example, in the minimal case $N' =1$ and $\Delta g =0$, one has $T_{\nu +{\rm D}} \simeq 0.815 \, T_{\nu}^{\rm SM}$.
Prior to rethermalisation, the extra radiation contribution is negligible. After rethermalisation and after the $\phi'$ phase transition,
occurring below $1\,{\rm MeV}$, one has
\be\label{extra}
\Delta N_\nu \simeq 3.043 \left[\, \left({3.043 + N' + 12/7 +4\Delta g /7 \over 3.043 + N' + 12/7 + 4\Delta g /7  - N_{\rm h}} \right)^{1\over 3} - 1 \right]  \,  .
\ee
In the minimal case one has $\D N_{\nu}\simeq 0.465$. This extra-radiation is produced after neutrino decoupling and, therefore,
it does not modify the prediction of the primordial helium abundance in standard BBN. However, if the phase transition occurs at temperatures
above nucleosynthesis, for $T \gtrsim T_{\rm nuc}\simeq 65\,{\rm keV}$, this extra-radiation modifies the primordial deuterium abundance.
Current constraints from deuterium on $\Delta N_{\nu}$ gives $\Delta N_{\nu}(t_{\rm nuc}) \lesssim 0.4$ at 95$\%$ C.L.  \cite{Pisanti:2020efz} and so there would be  a tension. This can be solved introducing extra-degrees of freedom in the dark sector, so that $\Delta g \neq 0$. However, using theoretical ab-initio energy dependencies  of nuclear rates, a group has recently obtained 
$\Delta N_{\nu} = 0.3 \pm 0.15$ \cite{Pitrou:2020etk},
hinting at the presence of non-standard physics. In this case the split majoron model would perfectly provide the modification of 
standard BBN able to solve this potential deuterium problem.  
It is intriguing that for phase transitions temperatures $T_\star' \sim 100\,{\rm keV}$
the peak of the GW spectrum produced during the phase transition from sound waves peaks 
exactly within the range of frequencies tested by 
NANOGrav. The signal is not sufficiently strong to explain the whole signal. However, at the peak and for 
values suficiently large, $\alpha \gtrsim 0.4$, the signal can be within the sensitivity of NANOGrav and, 
therefore,  if disentangled from the astrophysical
contribution expected from super massive black hole mergers, it could be revealed. As we mentioned,
for such large values of $\alpha$, the GW spectrum from sound waves is still poorly determined. 
Some recent calculations \cite{RoperPol:2023dzg} show that around the peak the GW spectrum can be 
strongly enhanced for $\alpha \gg 0.1$ with respect to the  expression Eq.~(\ref{omegasw}) derived from sound shell model and  recovered exactly for $\alpha \lesssim 0.1$. If the deuterium problem will be confirmed, then the search
of such contribution within the NANOGrav signal and a better understanding of the GW spectrum 
produced from first order phase transitions for $\alpha \sim 0.5$ 
will become crucial since, in combinations with the cosmological tensions, 
it would provide a clear signature of the split majoron model.
  
Finally, let us say that the effect of extra radiation at recombination now can be compensated by 
the modification of free streaming length of ordinary neutrinos in a way that positions of CMB peaks are unchanged. 
Actually, such a model has been shown to be able to give a better fit 
of cosmological observations  compared to the $\Lambda$CDM model \cite{Escudero:2019gvw,Sandner:2023ptm} 
and in this respect it is one of the best performing models in improving the 
fit of cosmological observations compared to the $\Lambda$CDM model \cite{Schoneberg:2021qvd}.
 
\section{Conclusions}

The generation of Majorana mass can lead to the production of a stochastic background of primordial GWs at the 
seesaw scale or scales, in the case of a multiple majoron model.  The split majoron model can modify pre-recombination era
in an interesting way, since it would address a potential emerging deuterium problem and ameliorate cosmological tensions
within $\Lambda$CDM. At the same time it can give a contribution to the NANOGrav signal.
\section*{Acknowledgments}

I wish to thank the organisers of the  Moriond 2024
Electroweak Interactions and  Unified Theories session for a very scientifically stimulating meeting. 
I also wish to thank S.F. King and M. Rahat for a fruitful collaboration \cite{paper1,paper2} 
on the topics discussed in my talk.
I acknowledge financial support from the STFC Consolidated Grant ST/T000775/1 and 
from the European Union's Horizon 2020 Research and Innovation
Programme under Marie Sk\l odowska-Curie grant agreement HIDDeN European
ITN project (H2020-MSCA-ITN-2019//860881-HIDDeN).

\end{document}